# Modelling and simulation of dependence structures in nonlife insurance with Bernstein copulas

Prof. Dr. Dietmar Pfeifer

Dept. of Mathematics, University of Oldenburg and
AON Benfield, Hamburg

Dr. Doreen Straßburger

Dept. of Mathematics, University of Oldenburg and
mgm consulting partners, Hamburg

Jörg Philipps

Dept. of Mathematics, University of Oldenburg

**Abstract:** In this paper we review Bernstein and grid-type copulas for arbitrary dimensions and general grid resolutions in connection with discrete random vectors possessing uniform margins. We further suggest a pragmatic way to fit the dependence structure of multivariate data to Bernstein copulas via grid-type copulas and empirical contingency tables. Finally, we discuss a Monte Carlo study for the simulation and PML estimation for aggregate dependent losses form observed windstorm and flooding data.

1. Introduction

The use of copulas for modelling and simulation purposes especially in nonlife insurance (internal models) has attained increasing interest in the recent years, see e.g. [3], Chapter 5 and the references given there. However, the discussion of potential copula models has so far mostly focussed on either the elliptical case (e.g. Gaussian and $t$-copula) or the Archimedian case (e.g. Gumbel-, Clayton-, Frank-copula and others). Although the use of Bernstein polynomials in one and more variables or in one and more dimensions (especially Bézier curves and surfaces) has a long tradition in numerical analysis and computer aided design, it seems that the true impact of Bernstein polynomials on copula models has been discovered only more recently, first in the framework of approximation theory (see e.g. [1], [5], [7], [8]) and later in particular in connection with applications in finance (see e.g. [2], [3], [10], [11]). Bernstein copulas possess several benefits compared to the traditional approaches:

- Bernstein copulas allow for a very flexible, non-parametric and essentially non-symmetric description of dependence structures also in higher dimensions
- Bernstein copulas approximate any given copula arbitrarily well
- Bernstein copula densities are given in an explicit form and can hence be easily used for Monte Carlo simulation studies.

In this paper, we take a special simple look on the construction of Bernstein copulas through discrete random vectors with uniform margins, and point out their connection to grid-type copulas discussed in [4] (also called *checkerboard copulas* in [2] and [5]). This view which can also be found in [2] and [5], however restricted to the bivariate case, opens a pragmatic approach to fit the dependence structure of observed data to Bernstein copulas via grid-type copulas and (multivariate) contingency tables. As an example, we present a Monte Carlo study on the aggregate risk distribution for dependent windstorm and flooding losses.

2. Some simple mathematical facts on Bernstein polynomials and copulas

**Lemma.** Let $B(m,k,z) = \binom{m}{k} z^k (1-z)^{m-k}$, $0 \leq z \leq 1$, $k = 0, \cdots, m \in \mathbb{N}$. Then we have

$$\int_0^1 m\, B(m-1, k, z)\, dz = 1 \quad \text{for } k = 0, \cdots, m-1.$$

Further,

$$\frac{d}{dz} B(m,k,z) = m \big[ B(m-1, k-1, z) - B(m-1, k, z) \big] \quad \text{for } k = 0, \cdots, m$$

with the convention $B(m-1, -1, z) = B(m-1, m, z) = 0$.

**Proof:**

$$\int_0^1 m\,B(m-1,k,z)\,dz = m\binom{m-1}{k}\mathrm{Beta}(k+1,m-k) = m\binom{m-1}{k}\frac{\Gamma(k+1)\Gamma(m-k)}{\Gamma(m+1)}$$

$$= \frac{m(m-1)!}{k!(m-k-1)!} \times \frac{k!(m-k-1)!}{m!} = 1.$$

Further, for $0 < k < m$,

$$\frac{d}{dz}B(m,k,z) = k\binom{m}{k}z^{k-1}(1-z)^{m-k} - (m-k)\binom{m}{k}z^{k}(1-z)^{m-k-1}$$

$$= m\binom{m-1}{k-1}z^{k-1}(1-z)^{(m-1)-(k-1)} - m\binom{m-1}{k}z^{k}(1-z)^{m-1-k}$$

$$= m\bigl[B(m-1,k-1,z) - B(m-1,k,z)\bigr]$$

which, by the above convention, also holds for $k \in \{0,m\}$. ◆

**Theorem.** For $d \in \mathbb{N}$ let $\mathbf{U} = (U_1,\cdots,U_d)$ be a random vector whose marginal component $U_i$ follows a discrete uniform distribution over $T_i := \{0,1,\cdots,m_i - 1\}$ with $m_i \in \mathbb{N}$, $i=1,\cdots,d$. Let further denote

$$p(k_1,\cdots,k_d) := P\left[\bigcap_{i=1}^{d}\{U_i = k_i\}\right] \text{ for all } (k_1,\cdots,k_d) \in \underset{i=1}{\overset{d}{\times}} T_i.$$

Then

$$c(u_1,\cdots,u_d) := \sum_{k_1=0}^{m_1-1}\cdots\sum_{k_d=0}^{m_d-1} p(k_1,\cdots,k_d)\prod_{i=1}^{d} m_i B(m_i-1,k_i,u_i),\ (u_1,\cdots,u_d) \in [0,1]^d$$

defines the density of a $d$-dimensional copula, called *Bernstein copula*. We call $c$ the Bernstein copula density induced by $\mathbf{U}$.

**Proof.** For fixed $1 \leq j \leq d$ we obtain, according to the Lemma above,

$$\int_0^1 c(u_1,\cdots,u_d)\,du_j = \sum_{k_1=0}^{m_1-1}\cdots\sum_{k_d=0}^{m_d-1} p(k_1,\cdots,k_d)\prod_{\substack{i=1 \\ i\neq j}}^{d} m_i B(m_i-1,k_i,u_i)\int_0^1 m_j B(m_j-1,k_j,u_j)\,du_j$$

$$= \sum_{k_1=0}^{m_1-1}\cdots\sum_{k_d=0}^{m_d-1} p(k_1,\cdots,k_d)\prod_{\substack{i=1 \\ i\neq j}}^{d} m_i B(m_i-1,k_i,u_i)$$

Modelling and simulation of dependence structures in nonlife insurance with Bernstein copulas

$$= \sum_{k_1=0}^{m_1-1} \cdots \sum_{k_{j-1}=0}^{m_{j-1}-1} \sum_{k_{j+1}=0}^{m_{j+1}-1} \cdots \sum_{k_d=0}^{m_d-1} \left( \sum_{k_j=0}^{m_j-1} p(k_1,\cdots,k_d) \right) \prod_{\substack{i=1 \\ i \neq j}}^{d} m_i B(m_i-1,k_i,u_i)$$

$$= \sum_{k_1=0}^{m_1-1} \cdots \sum_{k_{j-1}=0}^{m_{j-1}-1} \sum_{k_{j+1}=0}^{m_{j+1}-1} \cdots \sum_{k_d=0}^{m_d-1} P\left( \bigcap_{\substack{i=1 \\ i \neq j}}^{d} \{U_i = k_i\} \right) \prod_{\substack{i=1 \\ i \neq j}}^{d} m_i B(m_i-1,k_i,u_i)$$

$$=: c^{[1,\cdots,j-1,j+1,\cdots,d]}(u_1,\cdots,u_{j-1},u_{j+1},\cdots,u_d)$$

for $(u_1,\cdots,u_{j-1},u_{j+1},\cdots,u_d) \in [0,1]^{d-1}$, i.e. $c^{[1,\cdots,j-1,j+1,\cdots,d]}$ is also a Bernstein copula density, but of dimension $d-1$ instead of $d$. (Note that for $j=1$, the symbol $[1,\cdots,j-1,j+1,\cdots,d]$ reads $[2,\cdots,d]$, likewise for $j=d$, correspondingly for the vectors of variables.) Continuing integration according to the remaining variables except for the variable $u_r$ for fixed $1 \leq r \leq d$, we end up with

$$c^{[r]}(u_r) = \int_0^1 \cdots \int_0^1 c(u_1,\cdots,u_d)\,du_1\cdots du_{r-1}\,du_{r+1}\cdots du_d = \sum_{k_r=0}^{m_r-1} P(U_r = k_r) m_r B(m_r-1,k_r,u_r)$$

$$= \sum_{k_r=0}^{m_r-1} \frac{1}{m_r} m_r B(m_r-1,k_r,u_r) = \sum_{k_r=0}^{m_r-1} B(m_r-1,k_r,u_r) = \sum_{k=0}^{m_r-1} \binom{m_r-1}{k} u_r^k (1-u_r)^{m-k} = 1$$

for all $u_r \in [0,1]$ which proves that the $r$-th marginal density of $c$ is that of a continuous uniform distribution over $[0,1]$, for every $1 \leq r \leq d$. ◆

Note that the line of proof above shows that if $\mathbf{V} = (V_1,\cdots,V_d)$ is a random vector with joint Bernstein copula density $c$ as above, then also any partial random vector $(V_{i_1},\cdots,V_{i_n})$ with $n < d$ and $1 \leq i_1 < \cdots < i_n \leq d$ possesses a Bernstein copula density $c^{[i_1,\cdots,i_n]}$ given by

$$c^{[i_1,\cdots,i_n]}(u_{i_1},\cdots u_{i_n}) = \sum_{k_{i_1}=0}^{m_{i_1}-1} \cdots \sum_{k_{i_n}=0}^{m_{i_n}-1} P\left( \bigcap_{\ell=1}^{n} \{U_{i_\ell} = k_{i_\ell}\} \right) \prod_{\ell=1}^{n} m_{i_\ell} B(m_{i_\ell}-1,k_{i_\ell},u_{i_\ell}), \ (u_{i_1},\cdots,u_{i_n}) \in [0,1]^n.$$

This means that the Bernstein copula density $c^{[i_1,\cdots,i_n]}$ is just the density induced by the partial random vector $(U_{i_1},\cdots,U_{i_n})$.

By integration, we obtain the Bernstein copula $C$ induced by $\mathbf{U}$ as

$$C(x_1,\cdots,x_d) := \int_0^{x_d} \cdots \int_0^{x_1} c(u_1,\cdots,u_d)\,du_1\cdots du_d = \sum_{k_1=0}^{m_1} \cdots \sum_{k_d=0}^{m_d} P\left( \bigcap_{i=1}^{d} \{U_i < k_i\} \right) \prod_{i=1}^{d} B(m_i,k_i,x_i),$$

Modelling and simulation of dependence structures in nonlife insurance with Bernstein copulasfor $(x_1,\cdots,x_d)\in[0,1]^d$. This can be verified by partial differentiation of $C$, using the above Lemma, and some rearrangements in the summation:

$$\frac{\partial}{\partial x_r}C(x_1,\cdots,x_d)=\sum_{k_1=0}^{m_1}\cdots\sum_{k_d=0}^{m_d}P\left(\bigcap_{i=1}^{d}\{U_i<k_i\}\right)\prod_{\substack{i=1\\i\neq r}}^{d}B(m_i,k_i,x_i)m_r\left[B(m_r-1,k_r-1,x_r)-B(m_r-1,k_r,x_r)\right]$$

$$=\sum_{k_1=0}^{m_1}\cdots\sum_{k_d=0}^{m_d}P\left(\bigcap_{\substack{i=1\\i\neq r}}^{d}\{U_i<k_i\}\cap\{U_r=k_r\}\right)\prod_{\substack{i=1\\i\neq r}}^{d}B(m_i,k_i,x_i)m_r B(m_r-1,k_r,x_r)$$

which, by iteration, finally leads to

$$\frac{\partial^d}{\partial x_1\cdots\partial x_d}C(x_1,\cdots,x_d)=c(x_1,\cdots,x_d),\ (x_1,\cdots,x_d)\in[0,1]^d.$$

There is also a natural relationship between Bernstein and grid-type copulas as discussed in [2], [5] and [4]. We refer to a slightly more general setup here.

**Definition.** Under the assumptions of the above theorem define the intervals $I_{k_1,\cdots,k_d}:=\bigtimes_{j=1}^{d}\left[\frac{k_j}{m_j},\frac{k_j+1}{m_j}\right]$ for all possible choices $(k_1,\cdots,k_d)\in\bigtimes_{i=1}^{d}T_i$. Then the function

$$c^*:=\prod_{i=1}^{d}m_i\sum_{k_1=0}^{m_1-1}\cdots\sum_{k_d=0}^{m_d-1}p(k_1,\cdots,k_d)\mathbb{1}_{I_{k_1,\cdots,k_d}}$$

is the density of a *d*-dimensional copula, called *grid-type copula* induced by **U**. Here $\mathbb{1}_A$ denotes the indicator random variable of the set *A*, as usual.

A natural interpretation of this copula is as follows: a random vector $\mathbf{W}=(W_1,\cdots,W_d)$ follows a grid-type copula iff the conditional distribution fulfills the conditions

$$P^{\mathbf{W}}\left(\bullet\,|\,\mathbf{U}=(k_1,\cdots,k_d)\right)=\mathcal{U}\left(I_{k_1,\cdots,k_d}\right)\ \text{for all}\ (k_1,\cdots,k_d)\in\bigtimes_{i=1}^{d}T_i,$$

where $\mathcal{U}(B)$ denotes the continuous uniform distribution over a *d*-dimensional Borel set *B* with positive Lebesgue measure and

$$\mathbf{U}=(k_1,\cdots,k_d)\ \Leftrightarrow\ \mathbf{W}\in I_{k_1,\cdots,k_d}$$

(i.e., **U** denotes in some sense the "coordinates" of **W** w.r.t. the grid induced by the $I_{k_1,\cdots,k_d}$).

Hence the Bernstein copula induced by **U** can be regarded as a naturally smoothed version of the grid-type copula induced by **U**, replacing the indicator functions

$$\mathbb{1}_{I_{k_1,\cdots,k_d}}(u_1,\cdots,u_d) = \prod_{i=1}^{d} \mathbb{1}_{\left(\frac{k_i}{m_i},\frac{k_i+1}{m_i}\right]}(u_i) \text{ by the polynomials } \prod_{i=1}^{d} B(m_i-1,k_i,u_i),\ (u_1,\cdots,u_d) \in [0,1]^d.$$

**Example 1.** The following graphs show the smoothing effect in case $d = 1$.

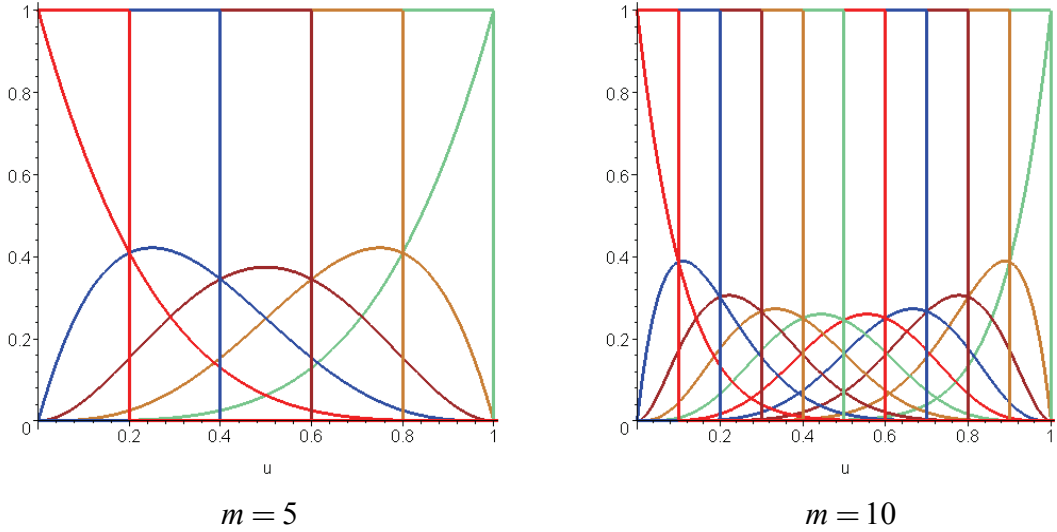

$m = 5 \qquad\qquad m = 10$

Natural generalizations of Bernstein and grid-type copulas are obtained if we look at suitable *partitions of unity*, i.e. families of non-negative functions $\{\phi(m,k,\bullet) | 0 \leq k \leq m-1,\ m \in \mathbb{N}\}$ defined on the unit interval $[0,1]$ with the following properties (see e.g. [5] or [7]):

- $\int_0^1 \phi(m,k,u)\,du = \dfrac{1}{m}$ for $k = 0,\cdots,m-1$

- $\sum_{k=0}^{m-1} \phi(m,k,\bullet) = 1$ for $m \in \mathbb{N}$.

In this case, a $d$-dimensional copula density $c^\phi$ induced by $\mathbf{U}$ is given by

$$c^\phi(u_1,\cdots,u_d) := \sum_{k_1=0}^{m_1-1}\cdots\sum_{k_d=0}^{m_d-1} P\left(\bigcap_{i=1}^{d}\{U_i = k_i\}\right)\prod_{i=1}^{d} m_i\phi(m_i,k_i,u_i),\ (u_1,\cdots,u_d) \in [0,1]^d.$$

The copula itself is accordingly given by

$$C^\phi(x_1,\cdots,x_d) = \sum_{k_1=0}^{m_1-1}\cdots\sum_{k_d=0}^{m_d-1} P\left(\bigcap_{i=1}^{d}\{U_i = k_i\}\right)\prod_{i=1}^{d} m_i\int_0^{x_i}\phi(m_i,k_i,u_i)\,du_i,\ (u_1,\cdots,u_d) \in [0,1]^d$$

(cf. [2] and [5] for the bivariate case with $m_1 = m_2$).

Note that

$$\phi(m,k,u) = B(m-1,k,u) = \binom{m-1}{k} u^k (1-u)^{m-1-k}$$

in case of Bernstein copulas and

$$\phi(m,k,u) = \mathbb{1}_{\left(\frac{k}{m},\frac{k+1}{m}\right]}(u)$$

for $0 \leq k \leq m-1$, $m \in \mathbb{N}$ in case of grid-type copulas.

Note further that any such family of functions $\{\phi(m,k,\bullet) | 0 \leq k \leq m-1, m \in \mathbb{N}\}$ induces immediately a new family $\{\phi_K(m,k,\bullet) | 0 \leq k \leq m-1, m \in \mathbb{N}\}$ for arbitrary, but fixed $K \in \mathbb{N}$ with similar properties via

$$\phi_K(m,k,\bullet) := \sum_{j=0}^{K-1} \phi(K \cdot m, K \cdot k + j, \bullet) \text{ for } k = 0,\cdots,m-1$$

since obviously

- $\int_0^1 \phi_K(m,k,u)\,du = \sum_{j=0}^{K-1} \int_0^1 \phi(K \cdot m, K \cdot k + j, u)\,du = \sum_{j=0}^{K-1} \frac{1}{K \cdot m} = \frac{1}{m}, \ k = 0,\cdots,m-1$

- $\sum_{k=0}^{m-1} \phi_K(m,k,\bullet) = \sum_{j=0}^{K-1} \sum_{k=0}^{m-1} \phi(K \cdot m, K \cdot k + j, \bullet) = \sum_{i=0}^{K \cdot m} \phi(K \cdot m, i, \bullet) = 1, \ m \in \mathbb{N}.$

For Bernstein copulas, this generalization has a direct impact on the smoothing effect pointed out in example 1. The following two graphs show this effect for $K = 3$ and $K = 10$. The case $K = 1$ is shown as a thin black line, for comparison.

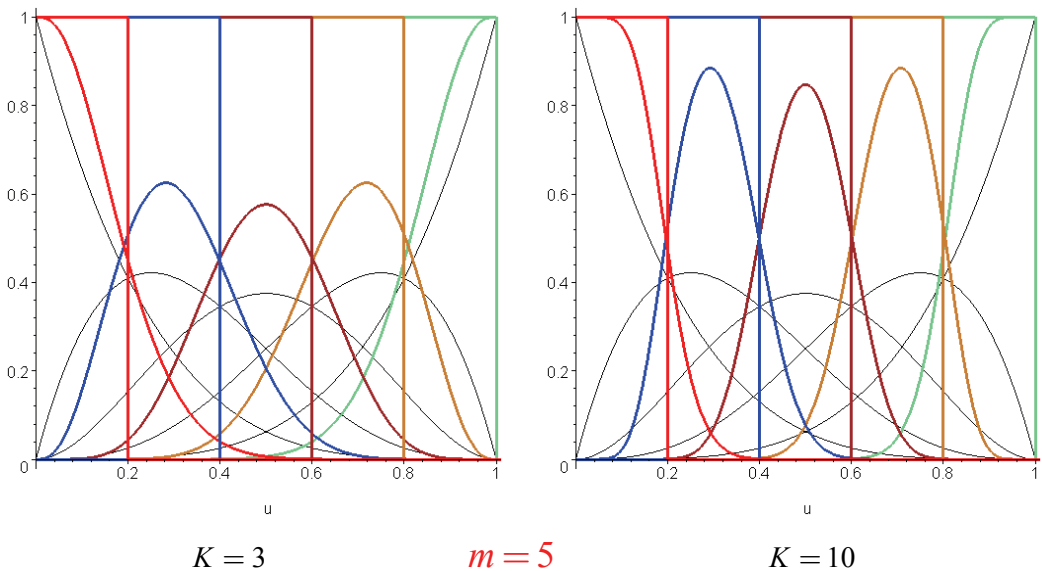

$K = 3$ $\quad\quad\quad m = 5 \quad\quad\quad$ $K = 10$

**Example 2.** Suppose that for dimension $d=2$ the joint distribution of $\mathbf{U}=(U_1,U_2)$ is given by the following table.

| $P(\mathbf{U}=(i,j))$ | | $i$ | | | |
|---|---|---|---|---|---|
| | | 0 | 1 | 2 | 3 |
| $j$ | 0 | 0,02 | 0 | 0,08 | 0,15 |
| | 1 | 0 | 0,03 | 0,12 | 0,10 |
| | 2 | 0,13 | 0,07 | 0,05 | 0 |
| | 3 | 0,10 | 0,15 | 0 | 0 |

The graphs below show jointly the grid-type and the Bernstein copula density induced by $\mathbf{U}$.

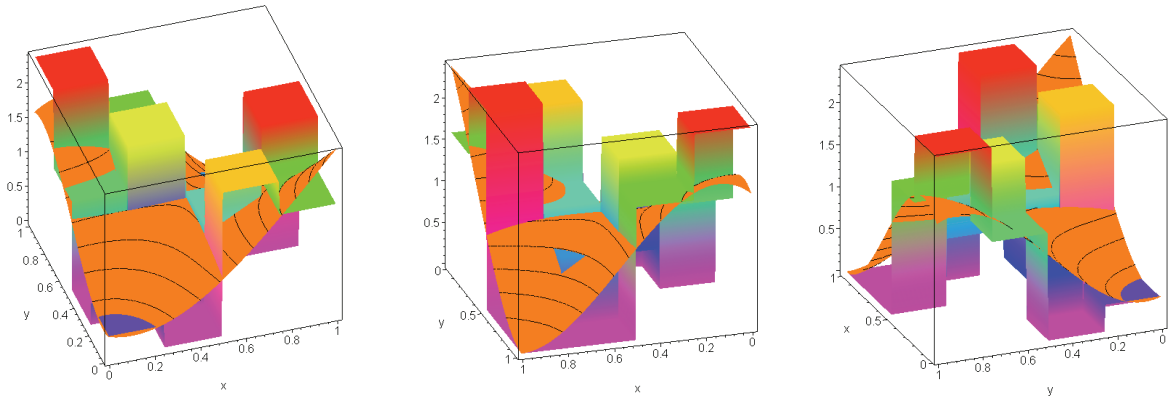

## 3. Fitting empirical data to grid-type and Bernstein copulas

In this section, we shall restrict ourselves to the case $d=2$, for simplicity. However, the method proposed here works accordingly in any dimension $d$.

Suppose that a bivariate data set of observations is given, for instance a time series of (economically adjusted) windstorm and flooding losses. One possible way to extract the dependence structure from the data is the *empirical copula scatterplot*, which is a plot of the joint relative ranks of the data. The following figure shows such a plot for a series of $n=34$ observation years.

Modelling and simulation of dependence structures in nonlife insurance with Bernstein copulas

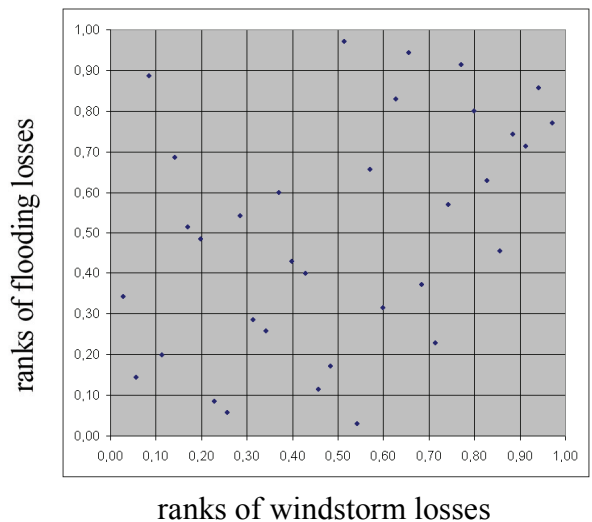

ranks of windstorm losses

In a first step, we want to fit these data to a grid-type copula with a given grid resolution, say $m_1 = m_2 = m = 10$. Counting the relative frequency of the data points in each of the $m_1 \times m_2 = 100$ target cells, we obtain the following contingency table $[a_{ij}]$ (matrix notation: $i$ = row index, $j$ = column index; rounded to 3 decimal places).

| upper cell boundary | 0,1 | 0,2 | 0,3 | 0,4 | 0,5 | 0,6 | 0,7 | 0,8 | 0,9 | 1,0 | sum |
|---|---|---|---|---|---|---|---|---|---|---|---|
| 1,0 | 0,000 | 0,000 | 0,000 | 0,000 | 0,000 | 0,029 | 0,029 | 0,029 | 0,000 | 0,000 | 0,009 |
| 0,9 | 0,029 | 0,000 | 0,000 | 0,000 | 0,000 | 0,000 | 0,029 | 0,000 | 0,000 | 0,029 | 0,009 |
| 0,8 | 0,000 | 0,000 | 0,000 | 0,000 | 0,000 | 0,000 | 0,000 | 0,029 | 0,029 | 0,059 | 0,012 |
| 0,7 | 0,000 | 0,029 | 0,000 | 0,000 | 0,000 | 0,029 | 0,000 | 0,000 | 0,029 | 0,000 | 0,009 |
| 0,6 | 0,000 | 0,029 | 0,029 | 0,029 | 0,000 | 0,000 | 0,000 | 0,029 | 0,000 | 0,000 | 0,012 |
| 0,5 | 0,000 | 0,029 | 0,000 | 0,029 | 0,000 | 0,000 | 0,000 | 0,000 | 0,029 | 0,000 | 0,009 |
| 0,4 | 0,029 | 0,000 | 0,000 | 0,000 | 0,029 | 0,029 | 0,029 | 0,000 | 0,000 | 0,000 | 0,012 |
| 0,3 | 0,000 | 0,000 | 0,000 | 0,059 | 0,000 | 0,000 | 0,000 | 0,029 | 0,000 | 0,000 | 0,009 |
| 0,2 | 0,029 | 0,029 | 0,000 | 0,000 | 0,059 | 0,000 | 0,000 | 0,000 | 0,000 | 0,000 | 0,012 |
| 0,1 | 0,000 | 0,000 | 0,059 | 0,000 | 0,000 | 0,029 | 0,000 | 0,000 | 0,000 | 0,000 | 0,009 |
| sum | 0,009 | 0,012 | 0,009 | 0,012 | 0,009 | 0,012 | 0,009 | 0,012 | 0,009 | 0,009 | |

Obviously, the observed marginal sums are not equal to $\frac{1}{m} = \frac{1}{10}$. We therefore consider the following optimization problem, in order to approximate the contingency table $[a_{ij}]$ by a uniform contingency table $[x_{ij}]$:

$$\min! \sum_{i=1}^{m}\sum_{j=1}^{m}(x_{ij} - a_{ij})^2 \text{ subject to}$$

$$\sum_{i=1}^{m} x_{ik} = \sum_{j=1}^{m} x_{\ell j} = \frac{1}{m} = \frac{1}{10} \text{ and } x_{\ell,k} \geq 0 \text{ for } k, \ell = 1, \cdots, m$$

Modelling and simulation of dependence structures in nonlife insurance with Bernstein copulas

The explicit solution of such a problem is in general not straightforward to find, although there exists a solution due to the *Karush-Kuhn-Tucker theorem* from optimization theory. Using a suitable software package like *octave* (a public domain computer algebra system), we obtain the following solution (rounded to 3 decimal places); see the code listing in the Appendix.

| upper cell boundary | 0,1 | 0,2 | 0,3 | 0,4 | 0,5 | 0,6 | 0,7 | 0,8 | 0,9 | 1,0 | sum |
|---:|---:|---:|---:|---:|---:|---:|---:|---:|---:|---:|---:|
| 1,0 | 0,003 | 0,000 | 0,002 | 0,000 | 0,003 | 0,027 | 0,032 | 0,027 | 0,003 | 0,003 | 0,1 |
| 0,9 | 0,032 | 0,000 | 0,001 | 0,000 | 0,002 | 0,000 | 0,031 | 0,000 | 0,002 | 0,031 | 0,1 |
| 0,8 | 0,000 | 0,000 | 0,000 | 0,000 | 0,000 | 0,000 | 0,000 | 0,020 | 0,025 | 0,055 | 0,1 |
| 0,7 | 0,003 | 0,027 | 0,002 | 0,000 | 0,003 | 0,027 | 0,003 | 0,000 | 0,032 | 0,003 | 0,1 |
| 0,6 | 0,000 | 0,025 | 0,029 | 0,021 | 0,000 | 0,000 | 0,000 | 0,025 | 0,000 | 0,000 | 0,1 |
| 0,5 | 0,003 | 0,028 | 0,002 | 0,025 | 0,003 | 0,000 | 0,003 | 0,000 | 0,032 | 0,003 | 0,1 |
| 0,4 | 0,027 | 0,000 | 0,000 | 0,000 | 0,027 | 0,021 | 0,026 | 0,000 | 0,000 | 0,000 | 0,1 |
| 0,3 | 0,003 | 0,000 | 0,002 | 0,054 | 0,003 | 0,000 | 0,003 | 0,028 | 0,003 | 0,003 | 0,1 |
| 0,2 | 0,025 | 0,020 | 0,000 | 0,000 | 0,055 | 0,000 | 0,000 | 0,000 | 0,000 | 0,000 | 0,1 |
| 0,1 | 0,003 | 0,000 | 0,061 | 0,000 | 0,003 | 0,026 | 0,002 | 0,000 | 0,002 | 0,002 | 0,1 |
| sum | 0,1 | 0,1 | 0,1 | 0,1 | 0,1 | 0,1 | 0,1 | 0,1 | 0,1 | 0,1 | |

A more pragmatic way to find at least a good suboptimal solution that can be easily implemented e.g. in spreadsheets is as follows. Consider the above optimization problem without the non-negativity conditions first. The equivalent Lagrange problem (which leads to a system of linear equations) is easy to solve and gives the (general) solution

$$x_{ij} = a_{ij} - \frac{a_{\bullet j}}{m} - \frac{a_{i\bullet}}{m} + \frac{2}{m^2} \quad \text{for} \quad i,j = 1,\cdots,m,$$

where the index • means summation, as usual. For the data set above, we thus obtain

| upper cell boundary | 0,1 | 0,2 | 0,3 | 0,4 | 0,5 | 0,6 | 0,7 | 0,8 | 0,9 | 1,0 | sum |
|---:|---:|---:|---:|---:|---:|---:|---:|---:|---:|---:|---:|
| 1,0 | 0,002 | -0,001 | 0,002 | -0,001 | 0,002 | 0,029 | 0,032 | 0,029 | 0,002 | 0,002 | 0,1 |
| 0,9 | 0,032 | -0,001 | 0,002 | -0,001 | 0,002 | -0,001 | 0,032 | -0,001 | 0,002 | 0,032 | 0,1 |
| 0,8 | -0,001 | -0,004 | -0,001 | -0,004 | -0,001 | -0,004 | -0,001 | 0,026 | 0,029 | 0,058 | 0,1 |
| 0,7 | 0,002 | 0,029 | 0,002 | -0,001 | 0,002 | 0,029 | 0,002 | -0,001 | 0,032 | 0,002 | 0,1 |
| 0,6 | -0,001 | 0,026 | 0,029 | 0,026 | -0,001 | -0,004 | -0,001 | 0,026 | -0,001 | -0,001 | 0,1 |
| 0,5 | 0,002 | 0,029 | 0,002 | 0,029 | 0,002 | -0,001 | 0,002 | -0,001 | 0,032 | 0,002 | 0,1 |
| 0,4 | 0,029 | -0,004 | -0,001 | -0,004 | 0,029 | 0,026 | 0,029 | -0,004 | -0,001 | -0,001 | 0,1 |
| 0,3 | 0,002 | -0,001 | 0,002 | 0,058 | 0,002 | -0,001 | 0,002 | 0,029 | 0,002 | 0,002 | 0,1 |
| 0,2 | 0,029 | 0,026 | -0,001 | -0,004 | 0,058 | -0,004 | -0,001 | -0,004 | -0,001 | -0,001 | 0,1 |
| 0,1 | 0,002 | -0,001 | 0,061 | -0,001 | 0,002 | 0,029 | 0,002 | -0,001 | 0,002 | 0,002 | 0,1 |
| sum | 0,1 | 0,1 | 0,1 | 0,1 | 0,1 | 0,1 | 0,1 | 0,1 | 0,1 | 0,1 | |

Seemingly, this "solution" is not feasible since it contains negative entries. A simple way to overcome this problem is a cell-wise additive correction with $a := -\min\{x_{ij} | 1 \leq i, j \leq m\}$ and consecutive norming; the final resulting contingency table $[y_{ij}] = \left[\dfrac{x_{ij} + a}{1 + m^2 \cdot a}\right]$ is given by

| upper cell boundary | 0,1 | 0,2 | 0,3 | 0,4 | 0,5 | 0,6 | 0,7 | 0,8 | 0,9 | 1,0 | sum |
|---|---|---|---|---|---|---|---|---|---|---|---|
| 1,0 | 0,004 | 0,002 | 0,004 | 0,002 | 0,004 | 0,024 | 0,026 | 0,024 | 0,004 | 0,004 | 0,1 |
| 0,9 | 0,026 | 0,002 | 0,004 | 0,002 | 0,004 | 0,002 | 0,026 | 0,002 | 0,004 | 0,026 | 0,1 |
| 0,8 | 0,002 | 0,000 | 0,002 | 0,000 | 0,002 | 0,000 | 0,002 | 0,022 | 0,024 | 0,046 | 0,1 |
| 0,7 | 0,004 | 0,024 | 0,004 | 0,002 | 0,004 | 0,024 | 0,004 | 0,002 | 0,026 | 0,004 | 0,1 |
| 0,6 | 0,002 | 0,022 | 0,024 | 0,022 | 0,002 | 0,000 | 0,002 | 0,022 | 0,002 | 0,002 | 0,1 |
| 0,5 | 0,004 | 0,024 | 0,004 | 0,024 | 0,004 | 0,002 | 0,004 | 0,002 | 0,026 | 0,004 | 0,1 |
| 0,4 | 0,024 | 0,000 | 0,002 | 0,000 | 0,024 | 0,022 | 0,024 | 0,000 | 0,002 | 0,002 | 0,1 |
| 0,3 | 0,004 | 0,002 | 0,004 | 0,046 | 0,004 | 0,002 | 0,004 | 0,024 | 0,004 | 0,004 | 0,1 |
| 0,2 | 0,024 | 0,022 | 0,002 | 0,000 | 0,046 | 0,000 | 0,002 | 0,000 | 0,002 | 0,002 | 0,1 |
| 0,1 | 0,004 | 0,002 | 0,048 | 0,002 | 0,004 | 0,024 | 0,004 | 0,002 | 0,004 | 0,004 | 0,1 |
| sum | 0,1 | 0,1 | 0,1 | 0,1 | 0,1 | 0,1 | 0,1 | 0,1 | 0,1 | 0,1 | |

Note that this matrix was used to feed the *octave* working sheet as an initial solution. The quadratic error between the contingency table $[y_{ij}]$ and the original $[a_{ij}]$ is 0,002175 while the quadratic error for the optimal solution is 0,000806 and hence somewhat smaller. For the remainder of the paper, we shall, however, use the contingency table $[y_{ij}]$, for simplicity; the optimal contingency table will produce mainly the same results here.

Note that for dimension $d > 2$, with the index sets $I^d := \{1, \cdots, m\}^d$ and, for $i \in \{1, \cdots, m\}$ and $k = 1, \cdots, d$, $I_k^d(i) := \{1, \cdots, m\}^{k-1} \times \{i\} \times \{1, \cdots, m\}^{d-k}$, the corresponding Lagrange optimization problem

$$\min! \sum_{(i_1 \cdots i_d) \in I^d} \left(x_{i_1 \cdots i_d} - a_{i_1 \cdots i_d}\right)^2 \text{ subject to}$$

$$x_{\bullet[k]}(i_k) := \sum_{(i_1 \cdots i_d) \in I_k^d(i_k)} x_{i_1 \cdots i_d} = \frac{1}{m} \text{ for } i_k \in \{1, \cdots, m\}, k = 1, \cdots, d \qquad (*)$$

has the solution

$$x_{i_1 \cdots i_d} = a_{i_1 \cdots i_d} - \frac{1}{m^{d-1}} \sum_{k=1}^{d} a_{\bullet[k]}(i_k) + \frac{d}{m^d} \text{ for } (i_1, \cdots, i_d) \in \{1, \cdots, m\}^d.$$

This can be seen as follows: putting the gradient of the Lagrange function

$$L = \sum_{(i_1\cdots i_d)\in I^d} \left(x_{i_1\cdots i_d} - a_{i_1\cdots i_d}\right)^2 + 2\sum_{k=1}^{d}\sum_{i_k=1}^{m} \lambda_{k,i_k}\left[x_{\bullet[k]}(i_k) - \frac{1}{m}\right]$$

to zero results in the $m^d$ additional equations (besides the side conditions (*))

$$\frac{\partial L}{\partial x_{i_1\cdots i_d}} = 2\left(x_{i_1\cdots i_d} - a_{i_1\cdots i_d}\right) + 2\sum_{k=1}^{d}\lambda_{k,i_k} = 0 \text{ for all } (i_1\cdots i_d)\in I^d. \qquad (**)$$

These two sets of equations are solved by

$$\lambda_{k,i_k} = \frac{a_{\bullet[k]}(i_k)}{m^{d-1}} - \frac{1}{m^d} \text{ and } x_{i_1\cdots i_d} = a_{i_1\cdots i_d} - \sum_{k=1}^{d}\lambda_{k,i_k} \text{ for } i_k \in \{1,\cdots,m\},\ k=1,\cdots,d,$$

which corresponds to the solution given above. Note that this solution fulfils (*) in particular because of

$$x_{\bullet[k]}(i_k) = a_{\bullet[k]}(i_k) - \frac{a_{\bullet[k]}(i_k)}{m^{d-1}}\cdot m^{d-1} - \frac{\sum_{(j_1\cdots j_d)\in I^d} a_{j_1\cdots j_d}}{m^{d-1}}\cdot (d-1)\cdot m^{d-2} + \frac{d}{m^d}\cdot m^{d-1} = \frac{d}{m} - \frac{d-1}{m} = \frac{1}{m}$$

for $i_k \in \{1,\cdots,m\},\ k=1,\cdots,d,$ as requested.

Note also that in the special case $d=2$, we have

$$a_{\bullet[1]}(i) = a_{i\bullet} \text{ and } a_{\bullet[2]}(j) = a_{\bullet j} \text{ for } i,j\in\{1,\cdots,m\}.$$

The above solution can be used as an initial solution for either the multidimensional Karush-Kuhn-Tucker approach or the simplified version described above, giving

$$y_{i_1\cdots i_d} = \frac{x_{i_1\cdots i_d}+a}{1+m^d a} \text{ with } a := -\min\left\{x_{i_1\cdots i_d} \mid 1 \leq i_1,\cdots,i_d \leq m\right\}.$$

The table $[y_{ij}]$ can be used to define the joint distribution of the discrete random vector $\mathbf{U} = (U_1, U_2)$ inducing the grid-type and Bernstein copulas, similar as in [2]. Note that in order to obtain a physically correct correspondence to the empirical copula scatterplot, we have to define

$$P(U_1 = i,\ U_2 = j) = y_{m-i, j+1} \text{ for } i, j = 0, \cdots, m-1.$$

The following graphs show the resulting copula densities.

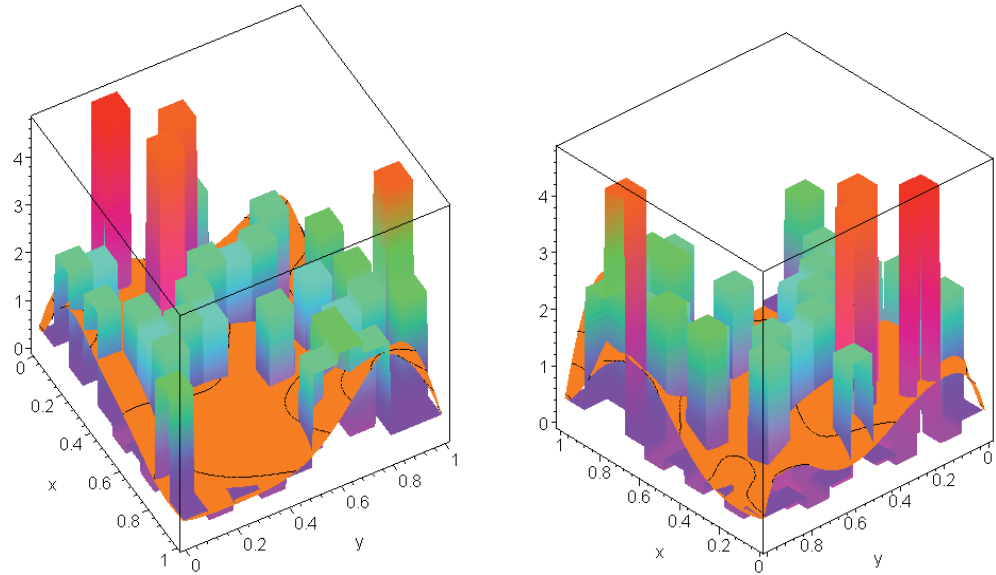

joint plot of the grid-type and Bernstein copula density induced by **U**

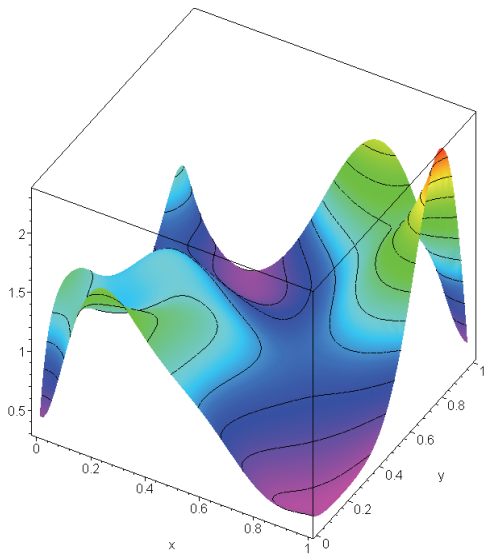
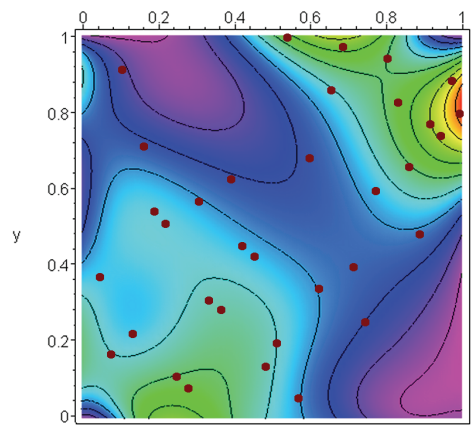

plot of the Bernstein copula density    contour plot of the Bernstein copula density, with original scatterplot superimposed

## 4. Simulating from Bernstein copulas

Since Bernstein copula densities are polynomials in $d$ variables, they are bounded over the unit cube $[0,1]^d$ by a constant $M > 0$, say which makes a stochastic simulation quite easy. The most convenient way is an application of the multivariate acceptance-rejection method (see e.g. [4], section 2.5.1):

- Step 1: generate $d+1$ independent uniformly distributed random numbers $u_1, \cdots, u_{d+1}$.
- Step 2: check whether $c(u_1, \cdots, u_d) > M u_{d+1}$. If so, go to Step 3, otherwise go to Step 1.
- Step 3: use $(u_1, \cdots, u_d)$ as a sample from the Bernstein copula.

The average rate of samples obtained by this procedure is $1/M$, as usual. Note that in our example, $M = 2,35$ is sufficient. From the 34 year time series of the logarithms of windstorm and flooding losses above the following marginal distributions were estimated, on the basis of a Q-Q-plot ($\mu =$ location parameter, $\sigma =$ scale parameter):

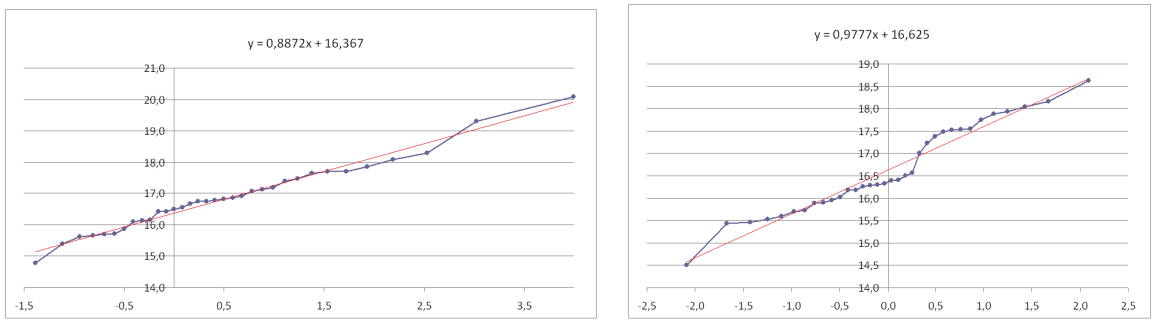

Q-Q-plots for log losses; left: windstorm, right: flooding

|  | Log windstorm losses | Log flooding losses |
| --- | --- | --- |
| Distribution | Gumbel | Normal |
| Parameters | $\mu = 16,367$ | $\mu = 16,625$ |
|  | $\sigma = 0,8872$ | $\sigma = 0,9777$ |

I.e., the windstorm losses are considered to be Fréchet distributed with extremal index $\alpha = 1/\sigma = 1,1271$ and the flooding losses are considered to follow a lognormal distribution. The following graphs show the results of a fourfold Monte Carlo simulation for the aggregate risk (windstorm and flooding) on the basis of 1000 pairs of points simulated from Bernstein copulas according to section 3 and the marginal distributions specified above. The four cases considered are:

- red line: Bernstein copula on the basis of a 4 x 4 grid (similar to section 3)
- green line: Bernstein copula on the basis of a 10 x 10 grid (exact data from section 3)
- blue line: independence case
- orange line: Gaussian copula estimated from original data

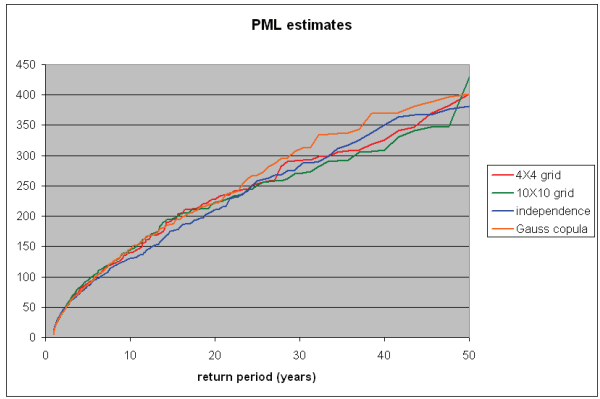
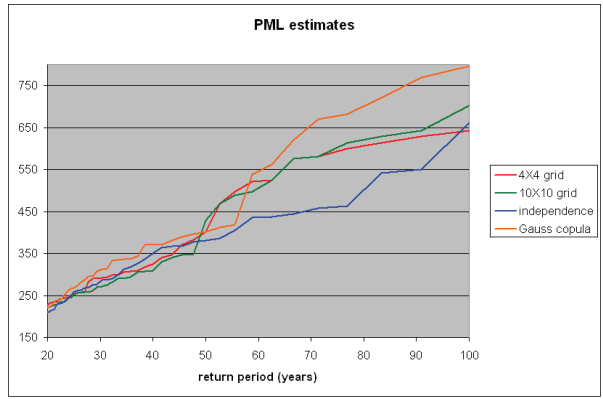
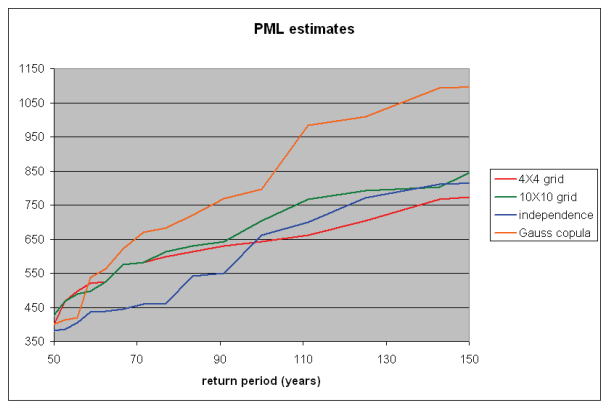
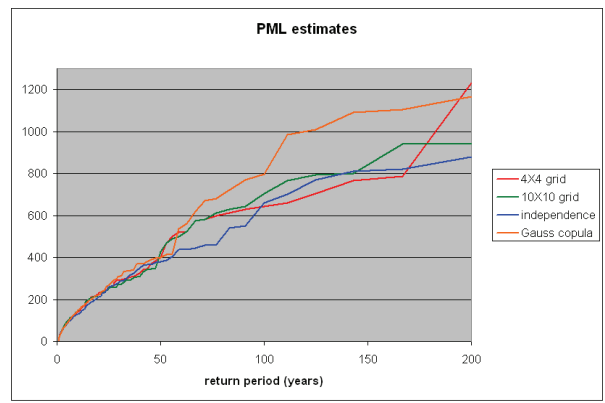

**Discussion:** All four copula models are quite close in the range of a return period of 50 years. Significant differences occur for higher return periods. It is interesting to observe that the PML estimates on the basis of Bernstein copulas lie between the independence case (lower bound) and the Gaussian copula (upper bound) in the range of 60 to 95 years return period.

The two Bernstein copula approaches are surprisingly close in the range up to a return period of 100 years, although the copula densities are clearly different.

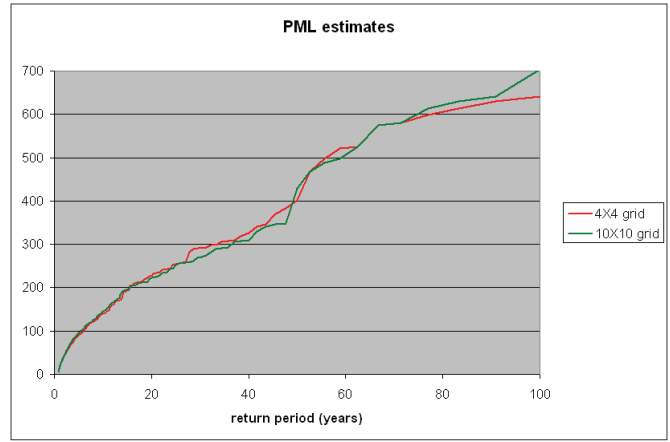

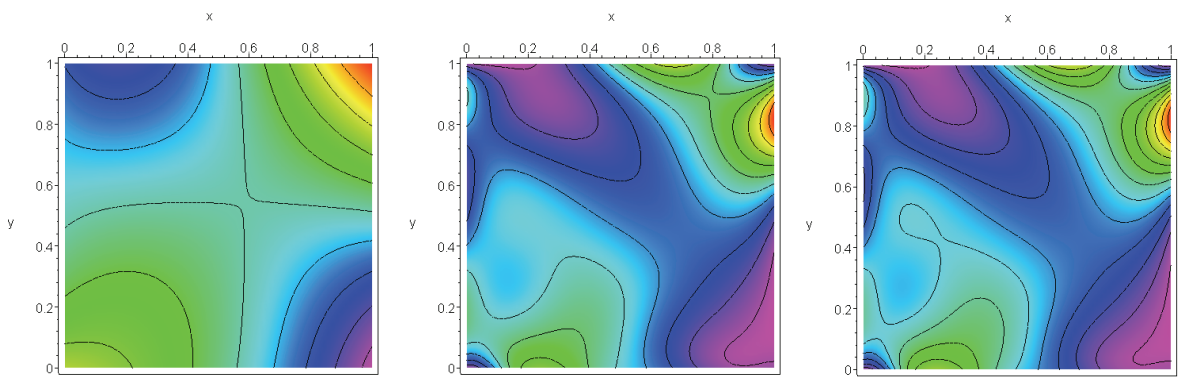

| 4x4 grid, approx. sol. | 10x10 grid, approx. sol. | 10x10 grid, opt. sol. |

contour plot of Bernstein copula densities

Note that the coarser grid produces a single peak of the density in the right top corner while the finer grid produces two distinct peaks there. This effect results in a substantially higher PML estimate for return periods above 170 years for the Bernstein copula on the 4x4 grid, even higher than the Gaussian copula above 200 years return period.

**Conclusion.** Using Bernstein or grid-type copulas for modelling dependence structures gives in general a better fit to local unsymmetries than other (classical) copulas can achieve, but a good compromise has to be found between the number of data points and the underlying grid resolution. Also, as is pointed out in [2], both types of copulas show a zero upper tail dependence since the densities are bounded. However, since Monte Carlo studies as performed here are always finite, this problem can be reduced for practical purposes by choosing a higher grid resolution.

**Acknowledgement.** We would like to thank Lena Reh for some stimulating discussions on the topic of copulas and pointing out some of the references to us.

Modelling and simulation of dependence structures in nonlife insurance with Bernstein copulas# Appendix

*octave* source code for the KKT-optimization problem from section 3

```
function x=bernsteinopt(A,x)
%A is the contingency table [a_ij] obtained by the data and x is the initial value.
%A possible initial value is the approximative solution presented in the paper

m=length(A);

%reshaping matrices to vectors
a=-vec(A);
X0=vec(x);

% positivitiy
lb=[1:m^2]'*0;

% equality constraint
b=[2:2*m]'*0+1/m;

% equality constraint column sum=1
B=[ ];
for i=1:m
        Bnew=[ ];
        for j=2:i
                Bnew=[Bnew,[1:m]*0];
        end
        Bnew=[Bnew,[1:m]*0+1];
        for j=i+1:m
                Bnew=[Bnew,[1:m]*0];
        end
B=[B;Bnew];
end

% equality constraint row sum=1
for i=2:m
        Bnew=[ ];
        C=[1:m]*0;
        C(i)=1;
        for j=1:m
                Bnew=[Bnew,C];
        end
B=[B;Bnew];
end

% octave quadratic optimization tool
[X, OBJ, INFO, LAMBDA] = qp (X0, eye(m^2), a, B, b, lb, [ ],
      [ ], eye(m^2), [ ]);

%reshaping vectors to matrices
x=reshape(X,m,m);
```